# Machine Learning Applications in Estimating Transformer Loss of Life

Alireza Majzoobi, Mohsen Mahoor, Amin Khodaei
Dept. of Electrical and Computer Engineering
University of Denver
Denver, CO, USA
Alireza.Majzoobi@du.edu, Mohsen.Mahoor@du.edu, Amin.Khodaei@du.edu

*Abstract*— Transformer life assessment and failure diagnostics have always been important problems for electric utility companies. Ambient temperature and load profile are the main factors which affect aging of the transformer insulation, and consequently, the transformer lifetime. The IEEE Std. C57.91-1995 provides a model for calculating the transformer loss of life based on ambient temperature and transformer's loading. In this paper, this standard is used to develop a data-driven static model for hourly estimation of the transformer loss of life. Among various machine learning methods for developing this static model, the Adaptive Network-Based Fuzzy Inference System (ANFIS) is selected. Numerical simulations demonstrate the effectiveness and the accuracy of the proposed ANFIS method compared with other relevant machine learning based methods to solve this problem.

## NOMENCLATURE

*Parameters:*

| | |
|---|---|
| $F_{AA}$ | Aging acceleration factor of insulation. |
| $F_{AA,n}$ | Aging acceleration factor for the temperature which exist during the time interval $\Delta t_n$. |
| $F_{EQA}$ | Equivalent aging factor for the total time period. |
| $K$ | Ratio of load to rated load. |
| $m/n$ | An empirically derived exponent used to calculate the variation of $\Delta\Theta_H/\Delta\Theta_{TO}$ with changes in load. |
| $Q$ | Number of test datasets. |
| $R$ | The ratio of load loss at rated load to no-load loss on the tap position to be studied. |
| $\Delta t_n$ | Time interval. |
| $\theta_H$ | Winding hottest-spot temperature (°C). |
| $\theta_A$ | Average ambient temperature during the load cycle to be studied (°C). |
| $\Delta\theta_H$ | Average ambient temperature during the load cycle to be studied (°C). |
| $\Delta\theta_{TO}$ | Top-oil rise over ambient temperature (°C). |

*Subscripts:*

| | |
|---|---|
| H | Winding hottest-spot |
| i/U | Initial/Ultimate |
| R | Rated |
| TO | Top oil |
| w | Winding |

## I. INTRODUCTION

ASSET MANAGEMENT has always been an important task performed by electric utility companies to ensure a reliable and secure operation of the power system. The goal of the asset management is to repair and upgrade power system components in an efficient and timely manner in a way that the probable component failures, and the consequent power outages, are minimized [1]. This topic is now more important than ever as the electricity infrastructure, mainly built in 1950s and 60s, is aging and consumers' expectations of a reliable and high quality service are at all-time-high values.

Under the broad topic of asset management, transformers play a key role and have attracted distinct attention, owing to two main reasons: dependence of network reliability on transformers, as transformer failures and unscheduled outages can potentially lead to unplanned power outage, and the significant investments required for transformers where transformer repair, upgrade, or change is time-consuming and costly for grid operators [2],[3]. Transformer asset management is carried out through various methods such as maintenance plans, condition monitoring, routine diagnostic, online monitoring, and condition based maintenance (CBM) [2],[4]. Since the transformer insulation has a higher probability of failure than other components of the transformer, the transformer life highly depends on its insulation condition. In addition, the aging of transformer mainly depends on its internal temperature, specifically at the hottest spot, which is governed by transformer loading and ambient temperature [5], [6]. Therefore, the load profile is the important factor on transformers aging which should be considered in transformer asset management. In [7], the effect of temperature and electrical stress have been experimentally studied via measuring important characteristics of insulation. The IEEE Std. C57.91-2011 Guide for Loading Mineral-Oil-Immersed Transformers, provides a method for calculation of distribution and power transformers loss of life [8]. The study in [5] has used this standard for estimating time to failure of transformer insulation through prediction of ambient temperature and load profile, based on the historical data. In [9], data quality control and data screening procedures are applied to transformer thermal model, proposed by the mentioned IEEE standard, in order to improve the reliability of the model. In [10], transformer loss of life is estimated in MATLAB and validated via comparison with experimental data. The effect of survival data on the accuracy of transformer statistical lifetime models is analyzed in [11] via Monte Carlo simulations where the results show that the accuracy of models can be improved by consideration of survival data. In [6], [12] and [13], the effect of electric vehicle on distribution assets is investigated and a method for smart charging of electric vehicles to manage distribution and transmission assets including transformers is proposed. An intelligent framework for condition monitoring of transformers via processing of historic data and obtained data from online measurements is proposed in [3].

In this paper, transformers loss of life is estimated via Adaptive Network-Based Fuzzy Inference System (ANFIS) which is an integration of Artificial Neural Networks (ANN) learning process and fuzzy inference system. ANNs are originally inspired by the biological structures of brains of humans and animals, which have extreme ability to solve complex problem in different disciplines [14]-[17]. Authors in [14] use fuzzy modeling to strategize asset management in transformers, where the improvement in remnant life and the rate of aging in power transformer are achieved with fuzzy model system. To predict top oil temperature in transformers, an artificial neural network is modeled in [15]. Load current and ambient temperature as two inputs in input layer and top oil temperature as one output in output layer are considered. In [16], a simple and accurate thermal model based on an evolutionary algorithm called genetic program is provided. The experimental data in [16] are derived from advanced metering infrastructure and to estimated transformer lifetime and accordingly determine the time of the transformer replacement.

The primary objective in this paper is to convert the dynamic model for transformer loss of life calculation to a static model, without losing accuracy, and accordingly apply a highly efficient machine learning model. Based on the IEEE Std. C57.91-2011, transformer loss of life is intricately formulated as a dynamic model since the degree of transformer insulation aging in each time interval depends on the load ratio of transformer in the current and previous time intervals. In accordance with the principle of parsimony, the simplest model that can explain and model a phenomenon is to be preferred. Providing a static model for transformer loss of life is the first contribution of this paper. Numerical simulations, to be carried out this paper, justify that the transformer loss of life can be precisely estimated using a static data-driven model.

Different types of data-driven methods, such as Multi-Layer Perceptron (MLP) network and Radial Basis Function (RBF) network are available to estimate a static system. However, as a second contribution in this research, ANFIS, as a data-driven static method is used to estimate transformer loss of life. Comparison between ANFIS and other machine learning methods clarifies the strengths of this method to solve the problem at hand. The proposed machine learning method can also be used to detect any transformer cooling system failure in the field as these failures reduce the transformer life.

The rest of the paper is organized as follows. Section II briefly describes the IEEE standard for calculation of the transformer loss of life. The ANFIS method is explained in Section III. Section IV presents numerical simulations and analyses to show the effectiveness of the proposed method, as well as merits over other existing machine learning methods. Conclusions are presented in Section V.

## II. TRANSFORMER LOSS OF LIFE CALCULATION BASED ON THE IEEE STANDARD

The criteria of 50% tensile strength of insulation was utilized for insulation lifetime estimation until 1950s, when AIEE Transformers Committee issued a report indicated that the chemical test measurement of degree of polymerization is a much better indication of cellulosic insulation mechanical characteristics than loss of tensile strength. Aging of transformer insulation is a function of temperature, moisture content, and oxygen content over the time. The amount of moisture and oxygen is controllable via transformer oil preservation system, but the temperature is a function of ambient temperature and operating conditions. As temperature distribution is not uniform in a transformer, the highest temperature spot (hottest-spot), which has the highest degree of aging on insulator, is considered for the loss of life calculations. The experiments and researches in late 1940s and 1950s showed that the transformer insulation aging follows a modification of Arrhenius' chemical reaction rate theory. The experimental equation (1) indicates the per unit life of transformers based on Arrhenius' chemical reaction rate theory.

$$\text{Per unit life} = A \exp(\frac{B}{\theta_H + 273}), \quad (1)$$

where $A$ and $B$ are empirical constants. $A$ is modified per unit constant which is computed based on selection of 110 °C as the temperature for "one per unit life" and equals to $9.8 \times 10^{-18}$. $B$ is the aging rate and depends on cellulose aging rates. Different experiments and standards declare various amounts for $B$ in the range of 11350 and 18000. The IEEE Std.C57.91-2011 considers 15000 as an appropriate value for $B$. Equation (1) is the basis for finding aging acceleration factor (AAF) for a given load and ambient temperature as in (2).

$$F_{AA} = \exp(\frac{15000}{383} - \frac{15000}{\theta_H + 273}). \quad (2)$$

This equation (2) is utilized to calculate the equivalent aging of the transformer (3) in a desired time period (one day, one month, one year, etc.)

$$F_{EQA} = \sum_{n=1}^{N} F_{AA_n} \Delta t_n \Big/ \sum_{n=1}^{N} \Delta t_n, \quad (3)$$

where $\Delta t_n$ is time interval, $n$ is the time interval index and $N$ is the total number of time intervals. The percentage of insulation loss of life is accordingly calculated as

$$LOL(\%) = \frac{F_{EQA} \times t \times 100}{\text{Normal insulation life}}. \quad (4)$$

The IEEE Std.C57.91-2011 mentions 180000 hours as the normal insulation lifetime for distribution transformers. It should also be noted that phrase "loss of life" commonly means "loss of insulation life", although "insulation" is frequently omitted. Calculation of all aforementioned equations depends on computing hottest-spot temperature which consists of three terms,

$$\theta_H = \theta_A + \Delta\theta_{TO} + \Delta\theta_H , \quad (5)$$

where, $\theta_A$ represents ambient temperature, $\Delta\theta_{TO}$ is top-oil rise over ambient temperature, and $\Delta\theta_H$ is the winding hottest-spot rise over top-oil temperature. Equations (6) and (7) define $\Delta\theta_{TO}$ and $\Delta\theta_H$, respectively.

$$\Delta\theta_{TO} = (\Delta\theta_{TO,U} - \Delta\theta_{TO,i})(1 - \exp(-\frac{1}{\tau_{TO}})) + \Delta\theta_{TO,i} , \quad (6)$$

$$\Delta\theta_H = (\Delta\theta_{H,U} - \Delta\theta_{H,i})(1 - \exp(-\frac{t}{\tau_w})) + \Delta\theta_{H,i} . \quad (7)$$

The initial and ultimate $\Delta\theta_{TO}$ and $\Delta\theta_H$ in (6) and (7) are calculated as

$$\Delta\theta_{TO,i} = \Delta\theta_{TO,R}(\frac{K_i^2 R + 1}{R+1})^n , \quad (8)$$

$$\Delta\theta_{TO,U} = \Delta\theta_{TO,R}(\frac{K_U^2 R + 1}{R+1})^n , \quad (9)$$

$$\Delta\theta_{H,i} = \Delta\theta_{H,R} K_i^{2m} , \quad (10)$$

$$\Delta\theta_{H,U} = \Delta\theta_{H,R} K_U^{2m} . \quad (11)$$

It should be noted that although $K_i$ and $K_U$ in (8)-(11) are respectively initial and ultimate values of transformer load ratio in each time interval, the value of $K_i$ in each time interval is equal to the value of $K_U$ at the end of previous time interval. Therefore, these formula offer a dynamic model for the calculation of the transformer loss of life. It should be mentioned that $m$ and $n$ vary between 0.8 and 1 based on the transformer cooling mode [8, Table 4], but the proposed method is independent of transformer cooling mode and could be applied to different modes by regenerating input datasets. More details of these equations can be found in [8].

### III. ADAPTIVE NETWORK-BASED FUZZY INFERENCE SYSTEM (ANFIS)

ANFIS is a type of ANN which is based on Takagi-Sugeno neuro-fuzzy models. ANFIS is a fuzzy model that is not only based on expert knowledge but also has learning from data, owing to integrating both neural networks and fuzzy inference, so it reaps the benefits of the learning process to fine-tune the parameters of the membership function in the fuzzy "if-then" rules. In other words, in ordere to describe the behavior of a system, ANFIS combines the structure of the neural networks, which deal with the implicit konwledge and the learning methods, with the explicit knowledge of the Fuzzy inference system. The unification of these two methods, i.e., neural networks and fuzzy inference system, yields enhancement in estimation performance, and offers a unique opportunity to solve complex problems. There is a meticulous similarity between ANFIS structure and feedforward neural networks. This adaptive network consists of nodes and direction links to connect nodes to each other. The main advantage of the ANFIS is that it takes the advantage of hybrid learning process to estimate the ANFIS parameters. The hybrid algorithm divided the learning process into two different and independent steps: (1) learning weights adaption (2) nonlinear membership functions adoption. Dividing the learning process into two steps not only decrease the algorithm complexity, but also makes the learning process to be more efficient. A typical ANFIS structure is shown in Fig. 1 [18], [19].

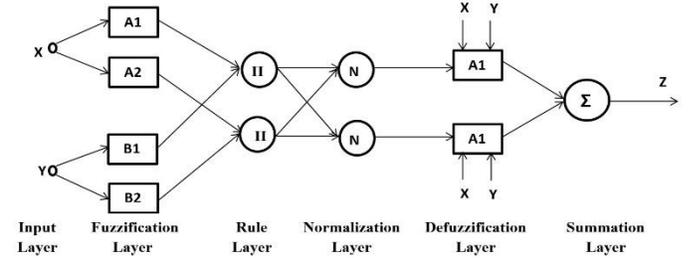

Figure 1. Schematic topology of ANFIS.

In order to evaluate and compare the performance and the accuracy of the proposed ANFIS model for transformer loss of life estimation, two performance measures are applied: Mean Square Error (MSE) and coefficient of determination ($R^2$) which are calculated in (12) and (13), respectively. $R^2$ ranges from 0 to 1, where $R^2=1$ means the proposed ANFIS model can estimate the actual transformer loss of life without error, and $R^2=0$ means the proposed ANFIS model cannot estimate the actual transform loss of life.

$$\text{MSE} = \frac{1}{Q}\sum_{q=1}^{Q}\left(Y_q - \hat{Y}_q\right)^2 , \quad (12)$$

$$R^2 = 1 - \left[\sum_{q=1}^{Q}\left(Y_q - \hat{Y}_q\right)^2 \bigg/ \sum_{q=1}^{Q}\left(Y_q - \bar{Y}\right)^2\right] , \quad (13)$$

In above equations, $Y_q$ is the actual output for the $q^{th}$ test dataset, $\hat{Y}_q$ is the estimated output for the $q^{th}$ test dataset and $\bar{Y}$ is the average of all actual outputs for test datasets.

It should be considered that data pre-processing is an important step in ensuring that bad data are detected and efficiently corrected before feeding to the proposed model.

### IV. SIMULATIONS

An hourly load profile of residential customers as well as the hourly temperature of a specific location in Chicago, IL [20] for one year are considered as input data to calculate the transformer loss of life based on the proposed method. The characteristics of the test transformer are borrowed from [5] and tabulated in Table I. Figs. 2, 3, and 4 show sample points for the hourly temperature of the test region in 2015, the ratio of load to nominal load of studied transformer over one year, and the calculated transformer loss of life, respectively.

TABLE I
CHARACTERISTICS OF STUDIED TRANSFORMER [5]

| $I_{rating}$ | R | m,n | $\Delta\theta_{H,R}$ | $\Delta\theta_{TO,R}$ | $\tau_{TO,R}$ |
|---|---|---|---|---|---|
| 934 A | 7.43 | 0.8 | 17.6 ºC | 53.9 ºC | 6.8 h |

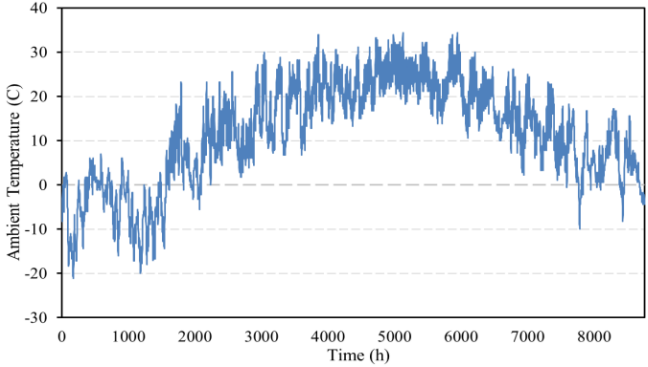

Figure 2. Ambient temperature of the test location [20].

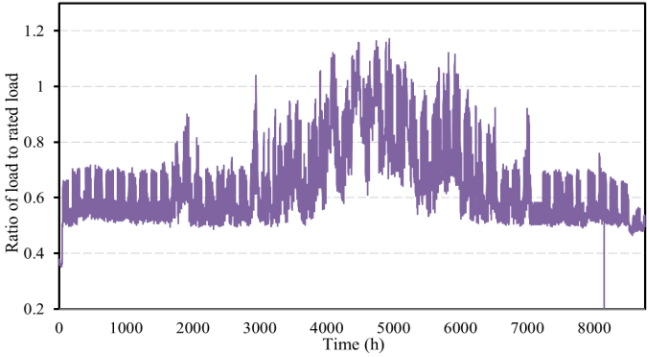

Figure 3. Hourly ratio of the actual load to the rated load of the transformer.

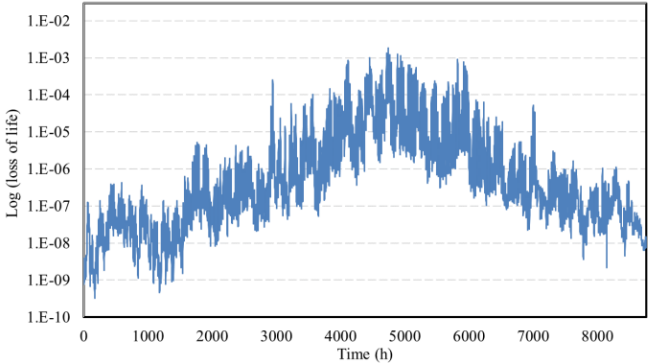

Figure 4. Hourly transformer loss of life, calculated based on the IEEE standard.

In order to prevent the overfitting problem, K-fold is used to check the performance. In this regard, the datasets are randomly split into k subsets; each time one of the k subsets is used as the test dataset and the reminder k-1 subsets are considered as training datasets. Then, the average error for all k trials is calculated to reach the best test and training datasets. Similar to the cross validation procedure, first each dataset is randomized to guarantee that training and test datasets are distributed fairly. Then, 30% and 70% of each dataset are randomly considered as the test datasets and the training datasets, respectively, in order to avoid the overfitting problem.

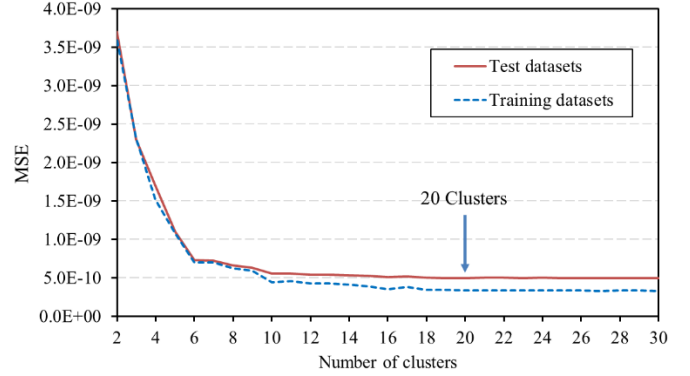

Figure 5. Specifying the number of clusters.

The proposed ANFIS method is modeled in MATLAB for hourly transformer loss of life estimation. The model uses fuzzy c-means clustering to specify the number of rules and membership functions in the Sugeno-type fuzzy inference structure. The number of clusters is specified based on the value of the MSE for the test and training datasets. This means that the MSE is plotted with only one epoch for the test and training datasets by changing the number of clusters. The point where does not have any decrement in the value of MSE is considered as a well-specified number of clusters for training the proposed ANFIS. Fig. 5 depicts the aforementioned procedure for specifying the number of clusters.

The generated ANFIS is trained by utilizing the input training datasets (current ambient temperature and load ratio), the output training datasets (calculated loss of life, based on the IEEE standard), 25 epochs, and 20 clusters. It should be noted that the optimum number of epochs, i.e. 25, is found based on trial and error. After the training, the test datasets are used to evaluate the estimated transformer loss of life. The calculated MSE and $R^2$ for the transformer loss of life in the test datasets are calculated as $2.946\times10^{-10}$ and 0.96, respectively. Fig. 6 compares the ANFIS-estimated loss of life with the actual loss of life, as well as the error which is defined as the difference between these two values. It is worth to mention that in order to be able to compare the estimated and the actual loss of life visually, Fig. 6 is plotted only for 100 samples of the test datasets.

In order to show the capability and the strength of the proposed ANFIS method to estimate the transformer loss of life, both MLP and RBF methods are further applied to estimate the transformer loss of life. For MLP modeling, a three-layered network based on error back propagation algorithm to train the MLP network in the MATLAB is used. Since there are two inputs (ambient temperature and load ratio) and one output (loss of life), two neurons for input layer and one neuron for output layer are considered for the generated network. On the basis of the rule of thumb, two neurons are employed in the hidden layer to process the MLP network. Hyperbolic tangent sigmoid transfer function is utilized between the input and the hidden layers, while linear transfer function is assigned from the hidden layer to the output layer.

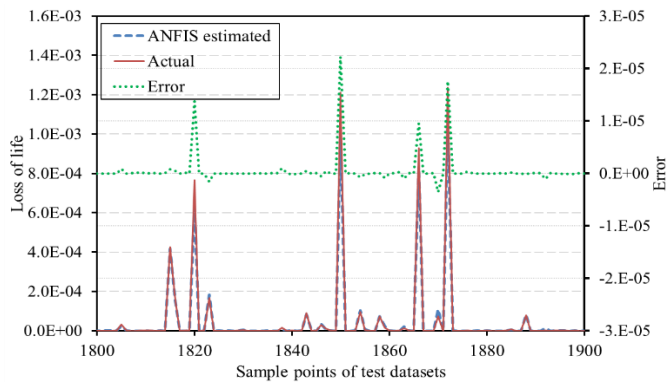
Figure 6. Comparison between the ANFIS estimated loss of life and calculated value based on the IEEE standard.

For RBF modeling, first, the calculated MSE value of the proposed ANFIS is set as the MSE goal for the RBF network. The simulation results indicate that the RBF network cannot reach the goal even with increasing the number of neurons. In other words, although the RBF network increases the number of neurons, it cannot improve the performance. Accordingly, the MSE goal for the RBF network is increased by 10% step sizes from 110% to 150% of the MSE value of the proposed ANFIS. The obtained results demonstrate that the RBF method reaches to 140% of the ANFIS MSE value by utilizing 1032 neurons. Table II compares the values of MSE, $R^2$ and computation time for these three methods. The proposed ANFIS method has the least MSE and the highest $R^2$ compared to the other two methods. However, computation time of the ANFIS method is slightly higher than other methods. The simulation is performed offline, so the computation time should not considered as a decisive factor in selecting the best method. Thus, taking all simulation results into account, it is confirmed that the ANFIS method has an outstanding capability to precisely estimate the transformer loss of life.

TABLE II
COMPARISON OF VARIOUS METHODS OF ANN FOR ESTIMATING TRANSFORMER LOSS OF LIFE

|  | MSE | $R^2$ | Computation time (s) | Rank |
|---|---|---|---|---|
| ANFIS | $2.946 \times 10^{-10}$ | 0.96 | 25.7 | 1 |
| MLP | $1.6027 \times 10^{-6}$ | 0.15 | 18.6 | 3 |
| RBF | $1.4 \times 2.946 \times 10^{-10}$ | 0.89 | 21.7 | 2 |

V. CONCLUSIONS

Transformers asset management has always been a critical subject for utility companies, due to their important role in power system reliability and significant capital and maintenance cost. The transformer lifetime highly depends on its insulation condition due to higher probability of insulation failure rather than other transformer components. This paper proposed a data-driven static model to estimate the transformer loss of life based on the IEEE Std. C57.91-2011. The ANFIS method was selected amongst different types of static methods than can be employed to solve this problem. The proposed ANFIS method was analyzed through numerical simulations, where it was shown that it can outperform other relevant methods in terms of ensuring the least MSE and the highest $R^2$.